\documentclass[a4paper,11pt]{article}
\usepackage{pos}
\usepackage{float}
\usepackage{hyperref} 
\usepackage{cleveref}
\crefname{figure}{Fig.}{Figs.}
\crefname{table}{Tab.}{Tabs.}
\crefname{equation}{Eq.}{Eqs.}
\crefname{section}{Sec.}{Secs.}
\usepackage{graphicx}
\usepackage{subcaption}
\usepackage{float} % for [H]
\usepackage{comment}

\title{Resolution Studies for Axion Searches\\with CUPID-0}
\ShortTitle{Resolution Studies for Axion Searches with CUPID-0}

\author*[a, b]{Livia Petrillo}
\onbehalf{on behalf of the CUPID-0 collaboration}

\affiliation[a]{Dipartimento di Fisica, Sapienza Università di Roma,\\P.le Aldo Moro 2, 00185, Roma, Italy}

\affiliation[b]{INFN, Sezione di Roma,\\P.le Aldo Moro 2, 00185, Roma, Italy}

\emailAdd{livia.petrillo@roma1.infn.it}

\FullConference{14th Young Researcher Meeting (14YRM2025)\\
30 September - 3 October 2025\\
Gran Sasso Science Institute, L'Aquila\\}

\begin{document}
\abstract{ABSTRACT\\Axions, hypothetical particles proposed to solve the strong CP problem and considered promising dark matter candidates, can be produced in the Sun and interact in detectors via couplings to photons, electrons, or nucleons. The CUORE and CUPID scintillating cryogenic calorimeters, originally developed to search for dark matter and neutrinoless double beta decay, are well suited to axion searches due to their excellent energy resolution, particle identification capability and low background. In order to search for high-energy solar axions at 5.5~MeV, a resolution study is carried out using calibration and background spectra, extrapolating the detector response in the hypothetical signal region. The analysis is based on exposures of 9.95~kg$\cdot$yr (CUPID-0 Phase~I) and 5.74~kg$\cdot$yr (CUPID-0 Phase~II). The energy resolution FWHM at 5.5~MeV is found to be $(39.8 \pm 2.1)$ keV for events fully contained in a single crystal, enabling axion searches with a background level lower than $10^{-3}$\,counts/(keV·kg·y).}
\maketitle

\vspace{-8pt}
\section*{Introduction}
\vspace{-8pt}
The Standard Model of particle physics is the most comprehensive and precise description of the fundamental constituents of matter and the forces that act between them. Although successful in many regards, with great ability in explaining and predicting a wide range of phenomena, this theory is far from complete. There are many questions that the Standard Model leaves unanswered: the overwhelming predominance of matter over antimatter in the Universe, the nature of dark matter, the origin of neutrino mass, the empirical absence of charge conjugation parity (CP) violation in strong interactions, and more. These open questions motivate new theoretical models that provide extensions to the Standard Model, and they inspire experimental efforts to test such theories, searching for new physics beyond the Standard Model.\\
Among such experiments is CUORE \cite{CUORE2025Science}, a cryogenic experiment located underground at the Gran Sasso laboratories; its upgrade CUPID \cite{CUPID2025cupid} is currently being developed with new and improved background rejection capabilities. The first demonstrator CUPID-0 collected data between 2017 and 2020 \cite{cupid0finalresults}.

\vspace{-8pt}
\section{The CUPID-0 experiment and the search for $\mathbf{0}\boldsymbol{\nu\beta\beta}$}
\vspace{-8pt}
The main goal of the CUORE and CUPID experiments is to investigate neutrino physics, aiming to answer a fundamental question about the nature of the neutrino: whether it is a Dirac particle, distinct from its antiparticle, or a Majorana particle, identical to it. \\The implications of this potential discovery are crucial: if neutrinos are Majorana, some processes that are forbidden by the Standard Model become possible and, in principle, should be observed. One of the most notable is neutrinoless double beta decay ($0 \nu \beta \beta$), an extremely rare hypothetical process whose observation would unambiguously prove the Majorana nature of neutrinos \cite{furry}. It consists of two simultaneous beta decays occurring within the same nucleus, with no neutrinos emitted in the final state. Compared to the Standard-Model-allowed two-neutrino double beta decay ($2 \nu \beta \beta$), neutrinoless double beta decay features no neutrinos in the final state and can be mediated by the exchange of virtual Majorana neutrinos between the weak vertices.

%\begin{figure}[h]
    %\centering
    %\includegraphics[width=0.5\linewidth]{Figures/scintillatingcryogeniccalorimeter.png}
    %\captionsetup{width=0.98\linewidth}
    %\caption{Schematic view of a scintillating cryogenic calorimeter, displaying the heat bath to ensure cryogenic temperatures, thermal sensor to measure heat and light detector to measure scintillation light.} 
    %\label{fig:dualreadout}
%\end{figure} 

In order to detect double beta decays, the experimental observable is the summed energy of the two emitted electrons; in case of neutrinoless double beta decay, a peak is expected in the summed-energy spectrum at the Q-value of the reaction. In CUPID-0, the energy is measured through an array of 26 ZnSe crystals positioned inside a cryostat to keep the apparatus at cryogenic temperatures ($\approx \hspace{-4pt} 10$ mK) \cite{firstresultscupid0, cupid0finalresults}. When an interaction takes place, a small amount of energy is deposited into the absorbing crystals and is partially converted into heat, while simultaneously emitting a light signal due to the scintillating properties of the crystals. The generated heat and light signals are simultaneously measured to determine the value of the deposited energy as well as the amount of emitted light corresponding to the event, allowing to reject background sources. This is the operating principle of a \textit{scintillating cryogenic calorimeter}.

Due to its excellent resolution and background rejection capabilities, this technology is well-suited to rare event searches of all kinds, extending well beyond neutrino physics. For instance, a detector of this kind allows to search for axions, hypothetical particles that contribute to the dark matter content of our Universe, which is the final goal of the procedure described in this work.

\begin{figure}
    \centering
    % First row
    \begin{subfigure}[t]{0.4\textwidth}
        \centering
        \includegraphics[width=\linewidth]{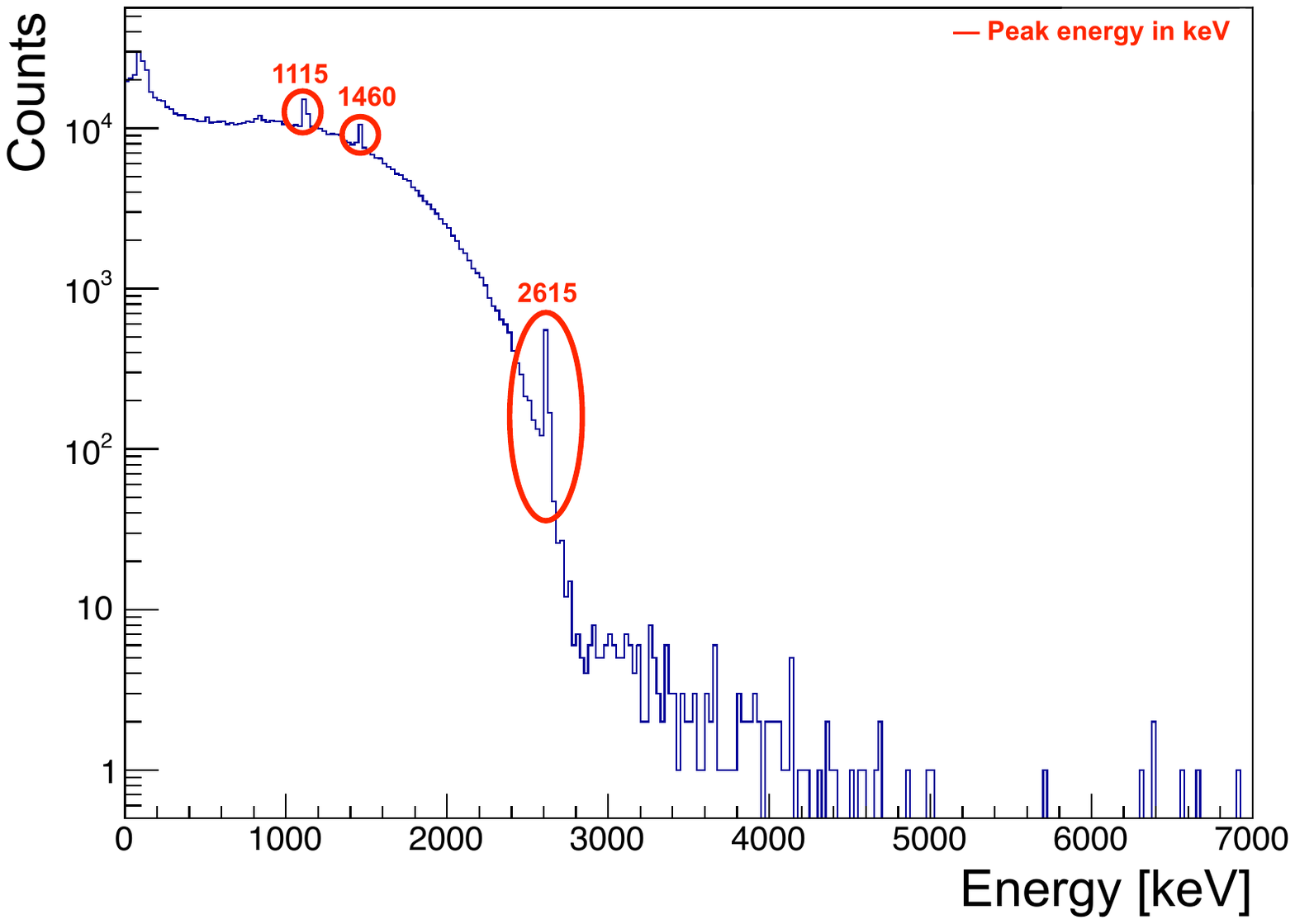}
        \caption{Background energy spectrum for $\beta/\gamma$ events.\vspace{-8pt}}
    \end{subfigure}
    \hfill
    \begin{subfigure}[t]{0.4\textwidth}
        \centering
        \includegraphics[width=\linewidth]{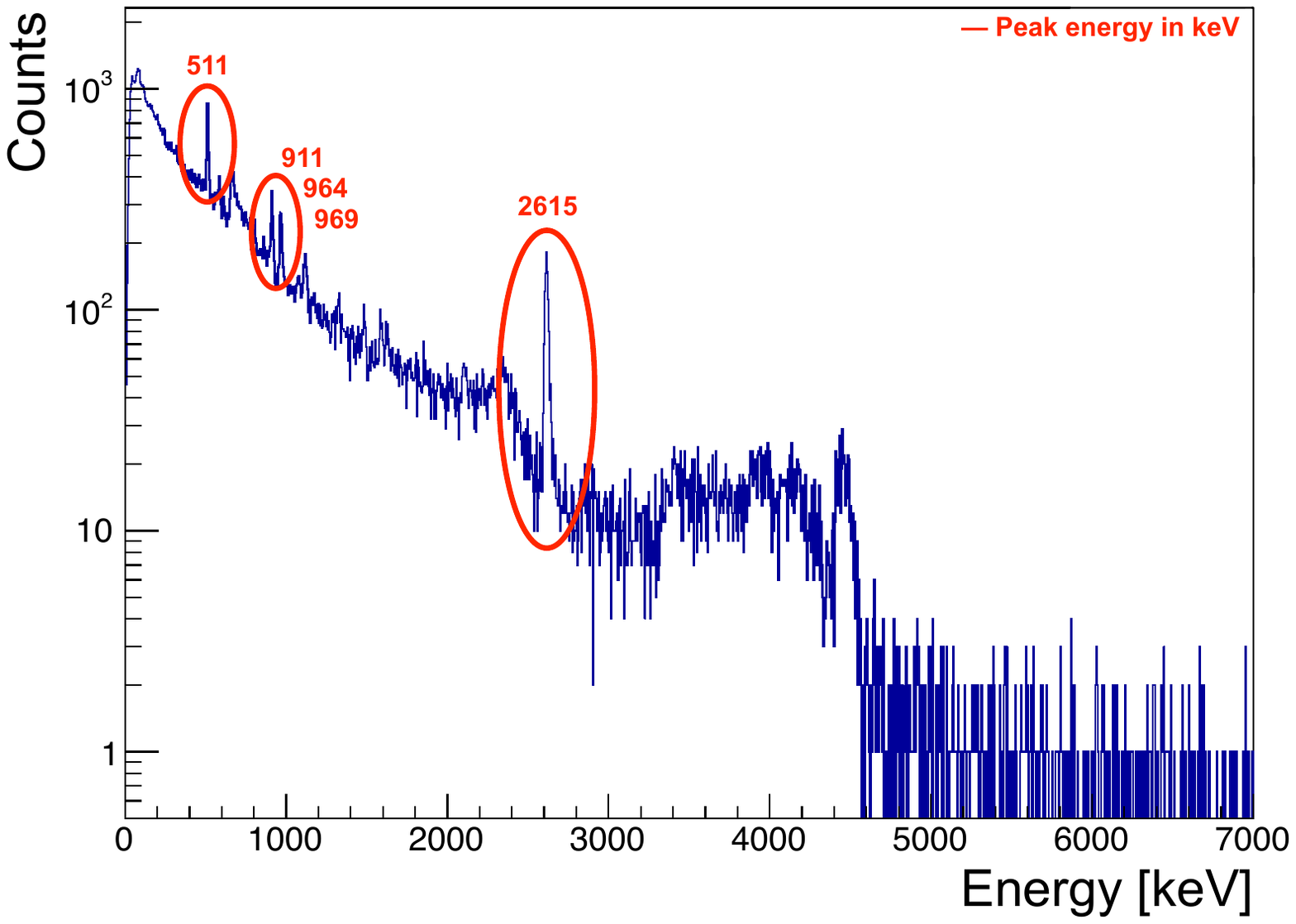}
        \caption{Calibration energy spectrum for $\beta/\gamma$ events.\vspace{-8pt}}
    \end{subfigure}
    
    \caption{Analyzed energy spectra of the $\beta/\gamma$ events collected by the CUPID-0 detector. The peaks selected for the resolution study are highlighted in red and labelled with the corresponding energies in keV. \vspace{-12pt}}
    \label{fig:four_images_bkg}
\end{figure}

\section{Detector calibration studies}
\vspace{-8pt}
Axions are hypothetical particles originally introduced to address some theoretical limitations of the Standard Model, related to the \textit{strong CP problem} \cite{pecceiquinn, diCortona2016}. Axions are expected to arise through different physical processes, both in laboratory settings and in astrophysical environments. Stars represent especially efficient sources thanks to their density and temperature, and particularly the Sun, due to its proximity, which may produce axions that reach the earth and are measured by detectors \cite{axioninducedpairproduction}.

The focus of this work will be on high-energy axions produced through nuclear reactions in the Sun, as their energies make them especially well-suited for detection with scintillating cryogenic calorimeters. High-energy solar axions are hypothetically emitted as a monochromatic 5.5 MeV flux, arising from axion–nucleon coupling in the $p + p \rightarrow \, ^3He + a$ reaction, which would occur as an alternative branch to the standard photon-producing channel.
Interacting with the CUPID-0 detector through different physical processes, axions give rise to characteristic signatures in the spectra. Axions can interact with matter through several processes whose relative importance depends on the underlying axion couplings, including the axioelectric effect, Compton-like scattering on electrons, and pair production. The resulting signatures display a peak at 5.5 MeV in the energy spectrum, matching the energy of the incident axions if fully absorbed. 
The CUPID-0  detector is optimized to operate at lower energies, up to about 3 MeV, where signals related to neutrinos are expected  \cite{cupid0finalresults}. This motivates the need to characterize the detector's response at high energy through a dedicated resolution study in order to set the grounds for an axion search.

The datasets considered for this analysis consist of background and calibration data from Phase I and Phase II of the CUPID-0 operation. 
The calibration data are collected in the presence of a $^{232}$Th source to characterize the detector's performance, while the background data are collected under normal operating conditions of the apparatus, without any additional source.
%In order to analyse the collected data, the recorded events are classified according to two criteria: the type of interaction and the event multiplicity. The first criterion, interaction type, distinguishes $\alpha$ interactions from $\beta/\gamma$ interactions. The second criterion, multiplicity, specifies whether the signal is detected by a single crystal or involves a coincident signal in two or more crystals.
%\noindent After classification, the energy spectra are produced by separating events according to multiplicity and particle type. Single-crystal events are labelled as {\small\textsf{M1}}, while events with multiplicity two are labelled as {\small\textsf{M2}}. For {\small\textsf{M1}} events only, particle identification is applied: $\alpha$ events are labelled {\small\textsf{a}}, while $\beta/\gamma$ events are grouped together and labelled {\small\textsf{b}}.\\
%\noindent For multiplicity greater than one, simultaneous energy depositions form a \textit{multiplet}. In this case, spectra can be constructed either using the energy deposited in a single crystal ({\small\textsf{M2}}) or using the sum of the energies of all crystals in the multiplet ({\small\textsf{M2sum}}).\\
%This procedure defines four spectra: \Ma, \Mb, \MM, and \MMsum. 
Considering both background and calibration data, the most prominent and well-defined peaks in the energy spectra are fitted individually to extract information about the detector’s response. The analysis procedure is performed on $\beta/\gamma$ events with multiplicity\footnote{The multiplicity of an event is defined as the number of crystals which detect a signal simultaneously (i.e. within the same time window of 20 ms) \cite{cupid0finalresults}. } M=1, as well as events with M=2; for the latter, the spectra are constructed by summing the energies of the two simultaneous events.\\ As an example, \cref{fig:four_images_bkg} shows the spectra produced by $\beta$ and $\gamma$ events with M=1, both with and without the calibration source, along with all the peaks analyzed in this work highlighted in red. Each selected peak is associated with a physical source whose corresponding nominal energy is obtained through an online database \cite{nudat}. The main lines in the spectra are produced by natural contaminants as well as isotopes generated in the decay chain of the calibration source: $^{40}$K, $^{208}$Tl, $^{65}$Zn, $^{228}$Ac, and more.

%The procedure is applied to both $\beta/\gamma$ events with multiplicity M=1 (only one crystal detects a signal at a given time) and for events with multiplicity M=2 (two crystals detect a signal in the same time window). 

\vspace{-4pt}
\subsection{Fit model}
\begin{figure}
    \centering
    \makebox[\textwidth]{%
        \includegraphics[width=0.7\textwidth]{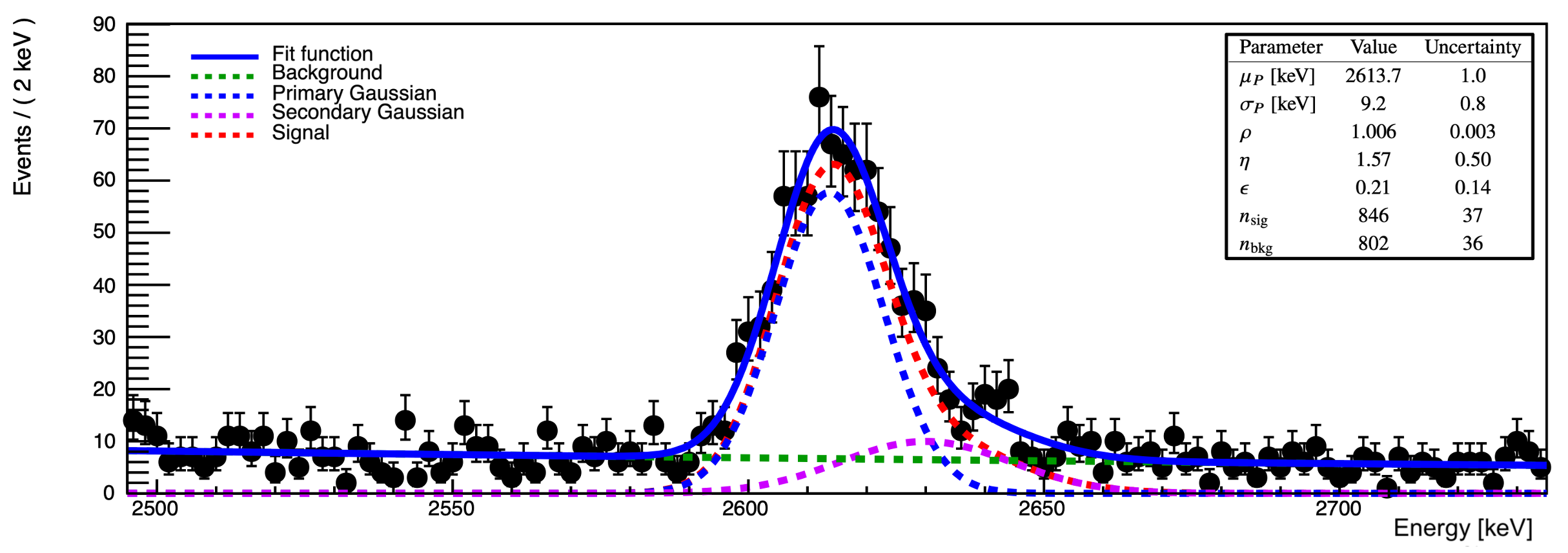}%
    }
    \caption{Fit of the 2615~keV $^{208}$Tl peak in the $\beta/\gamma$ calibration spectrum.  The best fit parameters are reported in the upper right box. \vspace{-12pt}}
    \label{fig:fit2615}
\end{figure}

\vspace{-4pt}
The selected peaks are fitted to extract their widths. To account for the slightly non-Gaussian detector response preciously observed in experiments like CUORE-0 \cite{searchfor0vbbcuore0} and CUORE, we adopt a multi-Gaussian model. This double-Gaussian approach is the simplest representation of the detector response across a broad energy range, likely arising from the combination of crystal modules with varying resolutions\footnote{The physical origin of this behavior is still under investigation. A possible explanation is that it is due to data from all crystal modules being combined into a single spectrum; some modules may operate sub-optimally and exhibit a worse resolution, producing broader peaks. Their contribution would then appear as a secondary, wider peak in addition to the primary peak associated with optimally performing detectors. The two peaks may not be centered in the same point, so the resulting distribution can show a slight asymmetry.}  \cite{searchfor0vbbcuore0}.\\
The fit function f(E) combines an exponential background and a signal component $\mathcal{G}$: 

\begin{equation}
    f(E) = \text{n}_{\text{bkg}} \cdot e^{-\lambda E} + \text{n}_{\text{sig}} \cdot \mathcal{G} \,(\mu_P, \sigma_P, \rho, \eta, \epsilon)
    \label{fitfunc}
\end{equation}
The response function $\mathcal{G}$ is the normalized sum of primary and secondary Gaussian components: 
\begin{equation}
    \mathcal{G} \,(\mu_P, \sigma_P, \rho, \eta, \epsilon) = (1-\epsilon) \cdot \mathsf{G_{primary}}(\mu_P, \sigma_P) + \epsilon \cdot \mathsf{G_{secondary}}(\rho \cdot \mu_P, \eta \cdot \sigma_P)
\end{equation}
where $\mathsf{G}\,(\mu, \sigma)$ denotes a Gaussian with mean $\mu$ and standard deviation $\sigma$.\\
To reduce the free parameters, the shape parameters ($\rho$, $\eta$, $\epsilon$) are first fixed using the high-statistics 2615 keV $^{208}$Tl calibration line (\cref{fig:four_images_bkg}(b)). The underlying assumption is that the detector should reproduce a similar response function across different energies, yielding consistent peak shapes. 
This approach was validated a posteriori by verifying that normalized residuals across all fits are centered at zero with unit standard deviation, indicating that the assumed response function provides an adequate description of the data across the explored energy range. While robust up to 2615 keV, the procedure is sensitive to potential deviations at higher energies when extrapolating at 5.5 MeV.\\
The 2615 keV peak is fitted using the model defined in \cref{fitfunc}. The resulting fit is shown in \cref{fig:fit2615}, where the $\beta/\gamma$ spectrum is displayed in the energy range
[2615-120, 2615+120] keV, along with the fit function and the best fit values of the fitted parameters.\\
\noindent After the fit is completed, the number of free parameters is reduced by fixing the parameters $\rho$, $\eta$ and $\epsilon$ to their best fit values based on the 2615 keV peak. The other selected peaks are then fitted using this reduced-parameter model.

\subsection{Extrapolation of the resolution}
\begin{figure}[H]
    \centering
    % First row
    \begin{subfigure}[t]{0.4\textwidth}
        \centering
        \includegraphics[width=\linewidth]{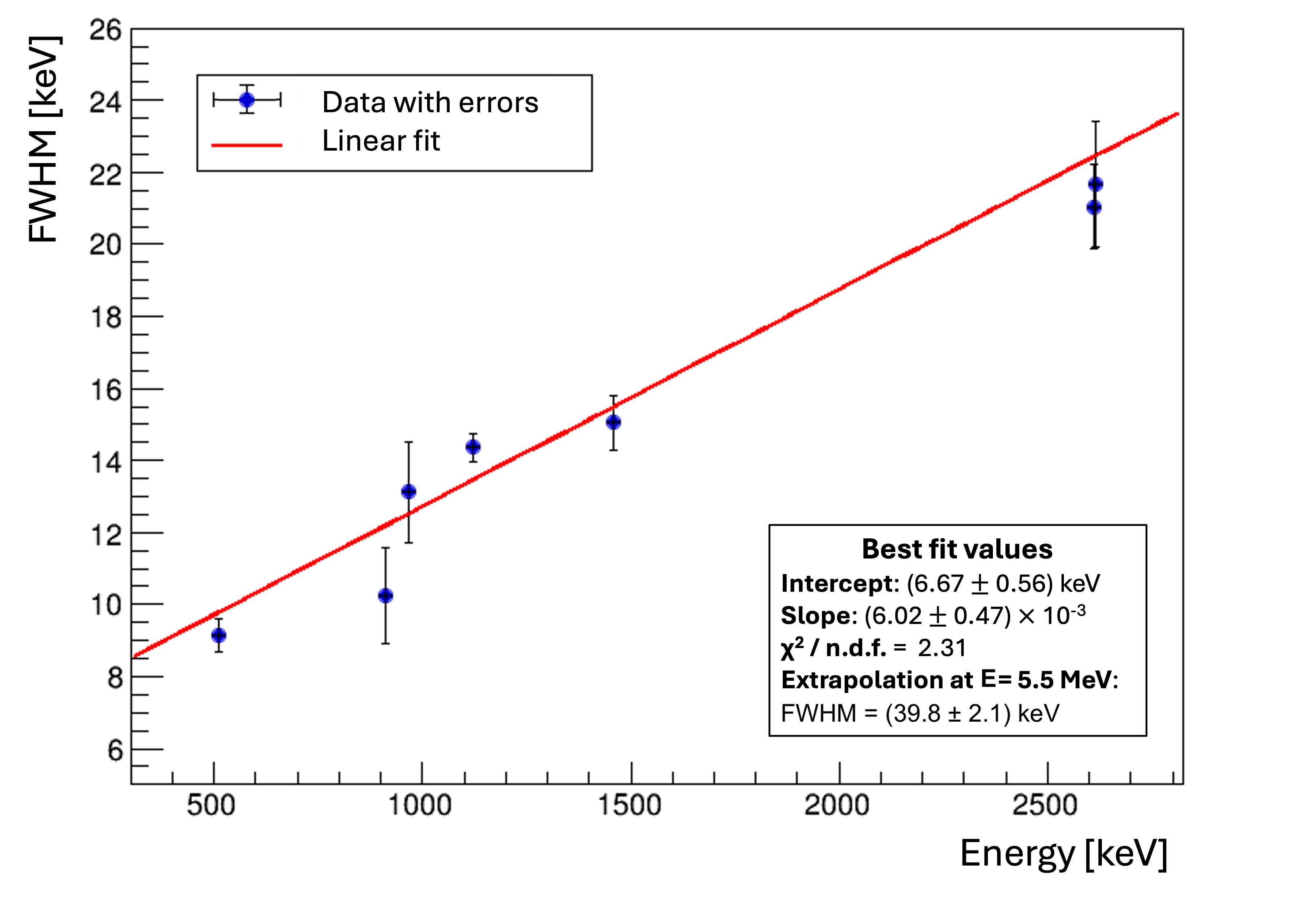}
        \caption{M=1 $\beta/\gamma$ events, background and calibration.\vspace{-8pt}}
    \end{subfigure}
    \hfill
    \begin{subfigure}[t]{0.4\textwidth}
        \centering
        \includegraphics[width=\linewidth]{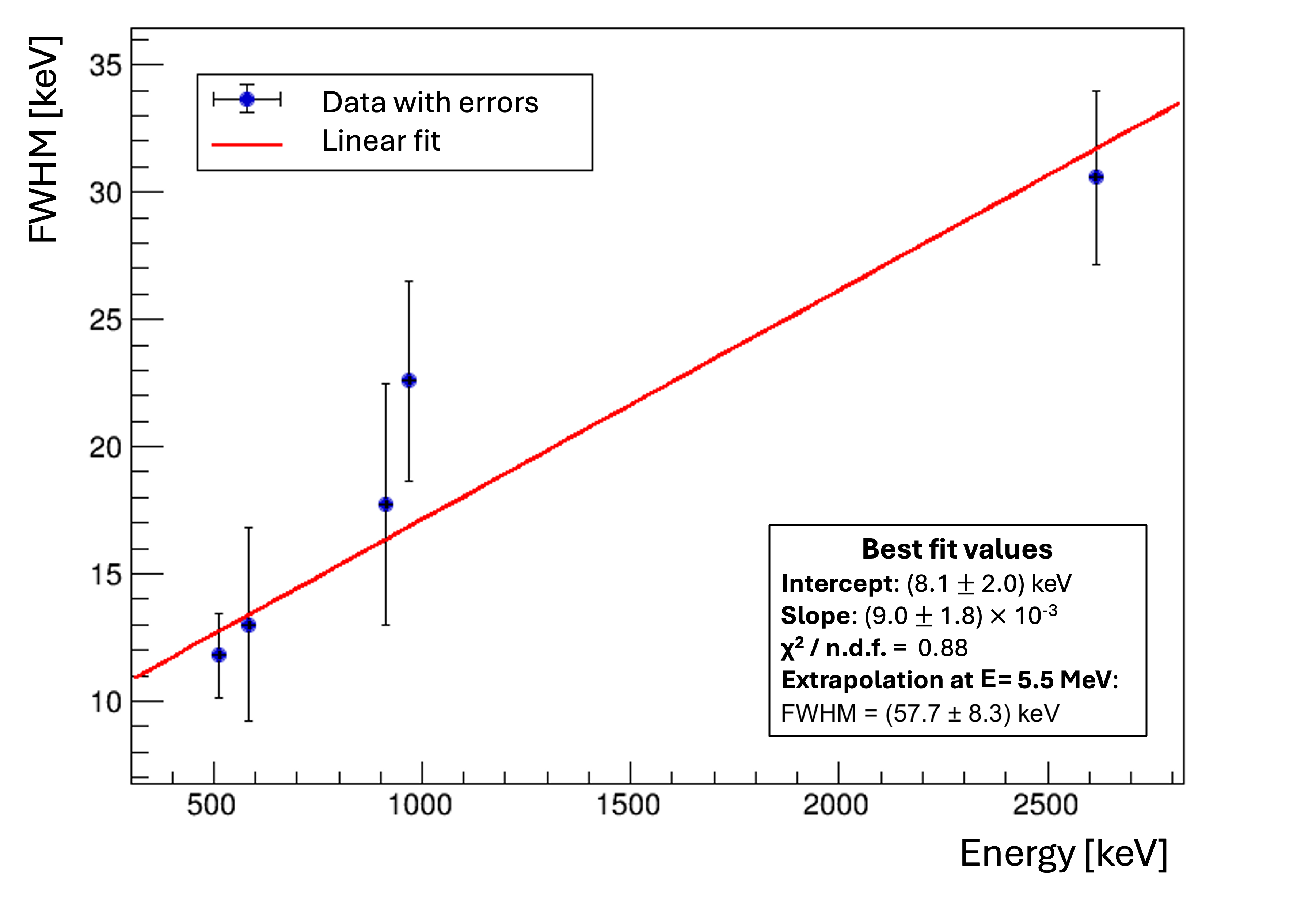}
        \caption{M=2 background events.\vspace{-8pt}}
    \end{subfigure}
    
    \caption{Linear fit of the energy resolution of the detector as a function of energy. \vspace{-12pt}}
    \label{fig:linearfits}
\end{figure}

After the fitting procedure of all the selected peaks is completed, the obtained parameters are used to extract the energy resolution of the detector at $5.5$~MeV, where a hypothetical monochromatic solar axion signal is expected. Determining the resolution at this energy is useful as it directly sets the expected width of such a signal in the spectrum. In order to extract the resolution, the full width at half maximum (FWHM = 2.335 $\sigma_P$) of each fitted peak is obtained from the Gaussian standard deviation $\sigma$ returned by the fit. By plotting the fitted FWHM as a function of the corresponding peak energy, one obtains the energy dependence of the detector resolution.
The resulting plots, as shown in \cref{fig:linearfits} separately for the M=1 and M=2 spectra, follow an increasing linear trend consistent with previous observations in CUPID-0 \cite{cupid0finalresults, cupid0finalresults_supplemental}. 
In standard calorimeter parameterizations, the energy resolution includes contributions from baseline noise, stochastic fluctuations, and a constant term accounting for non-statistical effects, including detector non-uniformities and calibration effects. In cryogenic bolometers such as CUPID-0, the stochastic term typically suppressed, so the resolution is dominated by an approximately energy-independent noise contribution at low energy and by an effective constant term at higher energies, leading to an approximately linear dependence of the FWHM on energy over the explored range.
Alternative parameterizations, such as a square-root dependence, were considered but are not statistically favored by the available data.

In the M=2 spectrum in \cref{fig:linearfits}(b), only the background data is shown, while the calibration data is discarded as it is affected by a systematic effect that increases the FWHM, as is expected in higher multiplicity spectra, which generally have a lower signal-to-noise ratio than single multiplicity spectra.
The effect originates from random, simultaneously occurring events: when the event rate is high, there is an increased probability of accidental coincidences, in which two unrelated detector signals are registered within the same coincidence window. When these random pairs of events are summed, their combined energy can accidentally fall near the energy of an existing peak, artificially increasing its width.
Calibration spectra are particularly affected by this phenomenon because the introduction of a calibration source introduces much higher statistics in the spectrum.\\
In contrast, in background runs the overall statistics is significantly lower, which suppresses the occurrence of accidental coincidences.
Consequently, the calibration and background data in the M=2 spectrum cannot be treated together as is done for M=1 data.

The obtained plots are then fitted with a straight line to extrapolate the detector resolution at higher energies, outside the directly measured range.
At 5.5 MeV, the extrapolated energy resolution for the M=1 $\beta/\gamma$ spectrum is $(39.8 \pm 2.1)$ keV, with the uncertainty being due to statistical effects from the linear fit.
The results are compatible with those obtained in Ref.\,\cite{cupid0finalresults} using $^{56}$Co, where the slope is found to be $(4.6 \pm 0.2) \times 10^{-3}$ (compatible within 3$\sigma$ with our results) and intercept $(6.23 \pm 0.24)$ keV (compatible with our results within 1$\sigma$). \\Similarly, for the M=2 background spectrum, the extrapolated resolution at 5.5 MeV is $(57.7 \pm 8.3)$ keV, consistent with sum in quadrature of the M=1 energy resolution.

\noindent The resulting FWHM of $\simeq 40$~keV at 5.5~MeV corresponds to a narrow axion signal window of a few tens of keV, leading to a negligible expected background given the observed level, below $10^{-3}$~counts/(keV$\cdot$kg$\cdot$y) as shown in \cref{fig:four_images_bkg}(a).
 This resolution study provides a quantitative understanding of the detector performance in the energy region of interest for 5.5 MeV solar axion searches, providing a foundation for defining the signal search strategy. Furthermore, the extremely low background level observed in this region demonstrates the excellent discovery potential of the experiment for high energy rare-event signals.

\end{document}